\documentstyle[12pt,epsfig]{article}
\newcommand{\be}{\begin{equation}}
\newcommand{\ee}{\end{equation}}
\newcommand{\ba}{\begin{eqnarray}}
\newcommand{\ea}{\end{eqnarray}}
\newcommand{\baa}{\begin{eqnarray*}}
\newcommand{\eaa}{\end{eqnarray*}}
\newcommand{\bb}{\begin{thebibliography}}
\newcommand{\eb}{\end{thebibliography}}

\newcommand{\lb}{\Lambda}
\newcommand{\ci}[1]{\cite{#1}}

\newcommand{\lab}[1]{\label{#1}}

\begin{document}
\begin{center}
{\large {\bf
Coulomb-hadron phase factor  and spin phenomena \\
in a wide region of transfer momenta  } } \\
\vspace{1cm}
 { S.B. Nurushev,
  \footnote{
   Institute for High Energy Physics, 142284 Protvino, Moscow
   }
  O.V.Selyugin
  \footnote{
   BLTP, JINR, 141980 Dubna, Russia; e-mail: selugin@thsun1.jinr.ru }
   M.N.Strikhanov$^{b}$  }
  \footnote{Moscow Engineering Physics Institute, Kashirrskoe Ave.31, 115409
   Moscow, Russia }

\end{center}

\begin{abstract}
  The Coulomb-hadron interference effects are examined at small and
  large $t$. The methods for the definition of spin-dependent
   parts of hadron scattering amplitude are presented.
  The additional contributions to analyzing power $A_N$
 and the double spin correlation parameter $A_{NN}$
  owing to the electromagnetic-hadron interference are determined
     in the diffraction dip domain
  of high-energy elastic hadron scattering.
\end{abstract}

     The energy region of RHIC in the fixed targed mode (FT)
 ($p_L= 30 \div 250 \ GeV/c$) at small transfer momenta
 is distinguished especially as the region where the second Reggeon
 contributions fall from an essential to an insignificant amount.
    The polarization phenomena determined at a low energy by
 the second Reggeon contributions also fall also very fast. Hence, in this
 domain the contributions into spin phenomena coming from
 the interference of the Coulomb and the hadron amplitudes
  play the important role \cite{nur}.
 On this background we must determine how fast fall in actually
 the spin-flip amplitudes of the second reggeons. Moreover, the high
 statistic and hence  small errors can allow us to find the
 contributions of the possible non-falling spin flip amplitudes
  originated from an instanton or high twist interactions.

      The definition of the structure of the high energy elastic
 hadron-hadron scattering amplitude at small angles is a topical
  problem \cite{selpl}. Such quantities as the total cross section and
 the phase of the scattering amplitude are relevant to
 the fundamental relations of the strong interaction theory as, for example,
 the dispersion relations.
 At present we cannot calculate these parameters  from the first principle
 or in the framework of the Perturbative Quantum Chromodynamics  (PQCD),
  therefore they should be measured   in experiment.

  The differential cross
  section and spin parameters $A_N$ and $A_{NN}$ are defined as
 \ba
  \frac{d\sigma}{dt}&=& \frac{2 \pi}{s^2}(|\phi_1|^2+|\phi_2|^2+|\phi_3|^2
   +|\phi_4|^2+4|\phi_5|^2),
								\lab{dsdt}
 \ea
 \ba
  A_N\frac{d\sigma}{dt}&=& -\frac{4\pi}{s^2}
                 Im[(\phi_1+\phi_2+\phi_3-\phi_4) \phi_5^{*})],  \lab{an}
\ea
and
 \ba
  A_{NN}\frac{d\sigma}{dt}&=& \frac{4\pi}{s^2}
                [ Re(\phi_1 \phi_2^{*} - \phi_3 \phi_4^{*})
                + 2 |\phi_5|^{2}],  \lab{ann}
\ea
 in framework of usual helicity representation.

   This information can be obtained from
  the precision measurement of the form of spin correlation parameters $A_N$
  and $A_{NN}$
  at small transfer momenta.  In this domain the analyzing power $A_N$
  is determined by the Coulomb-hadron interference effects. As we can
  calculate the Coulomb amplitude very precisely from the theory, we can
  obtain some information about the hadronic non-flip amplitude from this
  quantity.



   If we know the parameters of the hadron spin non-flip amplitude,
 the measurement of the analyzing power at small transfer momenta
 helps us to find the structure
 of the hadron spin-flip amplitude. For that  let us rewrite (\ref{an})
\ba
  A_N\frac{d\sigma}{dt}& =&
 2 [( ImF^{++}_h ReF^{+-}_c +ImF^{++}_c ReF^{+-}_c
        - ReF^{++}_h ImF^{+-}_c                          \\ \nonumber
   && -ReF^{++}_c ImF^{+-}_c )
     + (ImF^{++}_h ReF^{+-}_h -ReF^{++}_c ImF^{+-}_h    \\ \nonumber
     & &
        +  ImF^{++}_c ReF^{+-}_h -ReF^{++}_h ImF^{+-}_h )] \\ \nonumber
\ea
 where $F^{++}, F^{+-}, F^{--}$ are the spin-non flip, spin-flip and spin
 double-flip amplitudes, correspondingly, with ordinary connection
 with helicity amplitudes , for example,
 $F^{++}= (\phi_1 + \phi_3)/2$.
Let us denoted
$P^{'}$ is the part of the analyzing power which can be calculated if
   we know  $\sigma_{tot}, B, \rho$ for the non-flip amplitude.
   $\Delta P$ is the part depending on the hadron spin-flip amplitude.
\ba
  \Delta P = A_{N}^{exper.} -P^{'}.
\ea
Here $A_{N}^{exper.}$ is the measured value of the analyzing power.
Let us factor out the term  $F^{++}_c$ from the numerator,
the term  $(F^{++}_c)^2$ from the denominator and multiply
 $\Delta P$ by $F^{++}_c$ :
\ba
  A_{N}^{'}= (\Delta P ) F^{++}_c \simeq 2 \frac{-ImF^{+-}_h
      + ReF^{+-}_h (ImF^{++}_h/ReF^{++}_c)}
     {1 + (ImF^{++}_h/ReF^{++}_c)^2} .
\ea
  So, at the point $t$ where $ |ImF^{++}_h| = |ReF^{++}_c|$
  for the proton-proton scattering we have
\ba
   A_{N}^{'} = - Re F^{+-}_h - ImF^{+-}_h .
\ea
 And, as $t \rightarrow 0$
\ba
   A_{N}^{'} \rightarrow -2 ImF^{+-}_h
\ea
These calculations are compared with some model amplitude $F^{+-}$ on Fig.1.
at energies $\sqrt{s} = 50 $ and $500 \ GeV$ (short and long dashed line).

\begin{figure}
\centering
\mbox{\epsfysize=50mm\epsffile{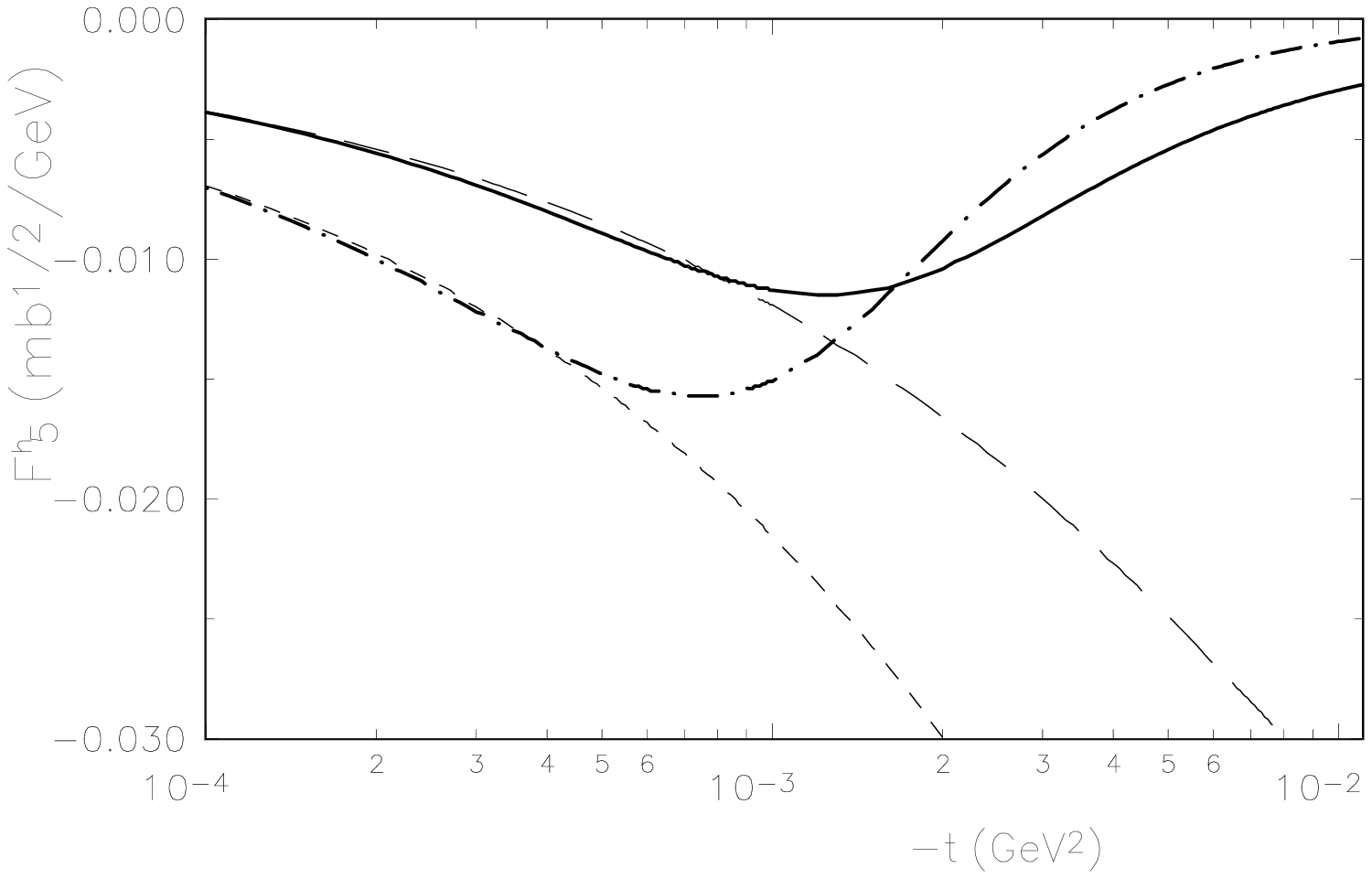}}

\vspace{-5mm}

\caption{The calculated $A_{N}^{'}/2$  }
\label{F1}
\end{figure}

    Now, let us examine the hadron double-spin-flip amplitude which
 has not the kinematical zero as $t \rightarrow 0$. This
 amplitude plays an important role at low energies, where they exist
 owing to a spin-spin interaction and an unnatural parity exchange,
 and has a sufficiently large magnitude.
        The measurement of  $A_{NN}$ at very small transfer momenta
  $\sim 10^{-4} \div 10^{-2} \ GeV^2$ gives information
  about the structure  of the double-spin flip amplitude
  and its energy dependence.
   For the double-spin correlation parameter $A_{NN}$ one can write:
\ba
A_{NN} (d\sigma/dt) &=&  2 F_{c}^{++} F_{c}^{--} + F_{c}^{++} Re(F_{h}^{--})
	    + \rho Im(F_{h}^{++}) Re(F_{h}^{--}) \\ \nonumber
      & &      + Im(F_{h}^{++}) Im(F_{h}^{--} +2 (F_{c}^{+-})^{2}
	    + 2 F_{c}^{+-} Re(F_{h}^{+-}) + 2 |F_{h}^{+-}|^2 \lab{anns}
\ea
Note, that
\ba
	     2 F_{c}^{++} F_{c}^{--} +2 (F_{c}^{+-})^{2} = 0,
\ea
 so these contributions, significant separately, disappear.

    If (\ref{ann}) is multiplied by the Coulomb amplitude, one can obtain
   in the range of small transfer momenta
\ba
  A_{NN}^{'}= A_{NN} F_{c}^{++} \simeq  \frac{
         Re(F_{h}^{--}) + Im(F_{h}^{--})  R
       +\rho [F_{c}^{--} + Re(F_{h}^{--})] R }
		     {1 + R^2},
\ea
 where $R= Im(F^{++}_{h}) / F_{c}^{++}$.
 So, at the point $t$, where  $|Im(F_{h}^{++})|=|F_{c}^{++}|$, it gives
\ba
  A_{NN}^{'} \simeq 0.5 [ Re(F_{h}^{--}) + Im(F_{h}^{--})].
\ea
 And as $t \rightarrow 0,$
\ba
  A_{NN}^{'} \rightarrow Re(F_{h}^{--})
\ea
These calculations are compared with some model amplitude $F_{h}^{--}$
 on Fig.2.
at energies $\sqrt{s} = 50$ and $500 \ GeV$ (short and long dashed line).

 Hence, the measurement of the double-spin correlation parameter at very
 small transfer momenta helps us to find the structure of the
  hadron double-spin-flip  amplitude.
   Note that to find the point where $|Im(F_{h}^{++})| = |Re(F_{c}^{++})|$,
 one  again needs know  exactly the real part of the spin-non-flip amplitude.

     So, we can conclude that the precise measurement
 of the spin correlation  parameters $A_N$  and $A_{NN}$ at very small
 transfer momenta $-t \sim 10^{-4} \div 10^{-2} GeV^2$ gives us a  way
 for the definition of the parameters of the high energy hadron interaction.
 It  allows us to reconstruct
 the structure of the high energy elastic scattering amplitude. Moreover,
 the precisely definition of the position of maximum of the analysing power
 permit us to define the value of the total cross section
 in an independent way.
  Hence, such a measurement is to be included in future research programs
  at HERA, RHIC and LHC.

\begin{figure}
\centering
\mbox{\epsfysize=75mm\epsffile{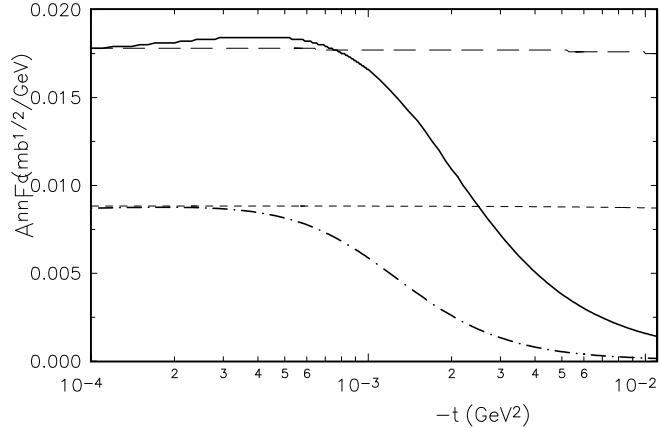}}
\caption{
The calculated $A_{NN}^{'}$ }
\label{F2}
\end{figure}

   At present, the spin effects owing to the Coulomb-nucleon
  interference (CNI) at very
 small transfer momenta are widely discussed in the aspect of  future
 spin experiments at RHIC and LHC.
 These effects are worse
 understood in the domain of the diffraction dip. In many respects, this
 is due to the fact that we do not know
 the Coulomb-hadron interference phase for
 not small transfer momenta and its impact on
 the magnitude of the spin effects.

    In papers \cite{bs,lap} the importance of CNI effects in the
 dip domain was pointed out. In \ci{bs} the polarization
  was calculated
  at sufficiently
 low energies with CNI-effect, but without taking accounted the phase
 of CNI and with grude approximation of the hadron non-flip spin amplitude.

    The phase  of the Coulomb-hadron interaction  has been calculated
and discussed  by many authors \cite{fi,isl,wy} and  has the form \cite{can}
\begin{eqnarray}
\varphi (s,t)&=&\mp [\gamma +\ln (B(s,t)|t| /2)
  +\ln (1 + 8/(B(s,t)\Lambda ^2)) +  \nu_c.         \label{fit}
\end{eqnarray}
where $B(s,t)$ is the slope of the nuclear amplitude;
$\Lambda$ is a constant entering into the dipole form factor and $\nu_c$
is an additional part coming from the pure Coulomb interaction
 and calculated
 in \ci{can} for small $t$.

  In \ci{selmpl}, the phase $\nu_c$ of the pure Coulomb amplitude
in the second Born approximation with the form factor in the monopole and
 dipole forms
has been calculated in a wide region of $t$.
It was shown that the behavior of $\nu_c$ at not small $t$ sharply differ
 from the behavior of $\nu_c$ obtained in \ci{can}.
  Let us calculate  the total phase factor  that can be used
 in the whole diffraction range of elastic hadron scattering.
   The scattering amplitude in an eikonal representation separating the Coulomb
  and the hadron eikonals can be write:
\ba
 F(s,t)= F_{c}(s,t) +F_{h}(s,t) [1
   - \frac{1}{2i F_{h}(s,t)} \int_{0}^{\infty} \chi_{c}(\rho)
       (1-e^{\chi_{h}(\rho,s)})  \rho d\rho J_0(\rho q) ]	\lab{fexp3}
\ea
  Taking into account the pure Coulomb phase $\varphi_{c}$
  we can rewrite (\ref{fexp3}) in  terms of the
 phase factor
\ba
 F(s,t)=
  F_{c} e^{i \alpha (\varphi_{ch} + \varphi_{c}) }  +F_{h}(s,t),
\ea
with
\ba
       \varphi_{c} = \ln{\frac{\lambda^2}{q^2}} \ + \ \nu.
\ea
 Here, $\lambda \rightarrow 0$ is the effective photon mass.
 Note that the divergent logarithmic term appears in the second
 Born approximation. Hence, if we take the Coulomb amplitude in first
 order, this term will be absent.

  To calculate of the Coulomb part in the Coulomb-hadron eikonal,
 let us expand the nucleon form-factor in  inverse powers of
 $\Lambda^2+q^2$ :
\ba
 G^2(q^2) & \equiv & [\frac{\Lambda^2}{\Lambda^2+q^2}]^4 \\ \nonumber
 &=& 1 - \frac{q^2}{\Lambda^2+q^2}
     - \frac{q^2 \Lambda^2}{(\Lambda^2+q^2)^2}
     - \frac{q^2 \Lambda^4}{(\Lambda^2+q^2)^3}
     - \frac{q^2 \Lambda^6}{(\Lambda^2+q^2)^4}
\ea
   Then, we obtain for the Coulomb part of the Coulomb-hadron
 eikonal:
\ba
 \chi_c(\rho) &=& -2i[\alpha( K_0(\rho\lambda) - K_0(\rho \Lambda)
   - \frac{\lb \rho}{2 \Gamma(2)} K_{-1}(\rho \Lambda)
   - \frac{\lb^2 \rho^2}{4 \Gamma(3)} K_{-2}(\rho \Lambda) \\ \nonumber
  &&  - \frac{\lb^3 \rho^3}{8 \Gamma(4)} K_{-3}(\rho \Lambda)]. \lab{ce1}
\ea
Here $K_{i}$ are  MacDonald functions and $\Gamma(n)$ is a gamma function.
  The first term gives the eikonal without the nucleon form-factor
 and others give the corrections that play an important role when $t$ differs
 from zero.
   Using the logarithmic representation of  $K_{0}$ we can
  take out from integral the first two terms, containing
  a divergences part and constant.  The rest part of these
  two first terms in numerator, containing only the hadron eikonal,
  are canceled by  the denominator.
 In result, the term with
 $\lambda$ is  reduced
 with the same term of $\chi_c$ at calculation observable
 and
  the total phase factor is:
\ba
 \varphi(s,t) = \ln{\frac{q^2}{4}} +2\gamma +\frac{1}{F_h(s,q)}
  \int_{0}^{\infty} \tilde{\chi}_{c}(\rho)
  (1 - \exp(\chi_h(\rho,s))J_{0}(\rho,q)d\rho , \lab{fei2}
\ea
 with
\ba
  \tilde{\chi}_c(\rho) = 2\rho \ln{\rho} +2\rho K_{0}(\rho \lb)
  [1+ \frac{5}{24} \lb^2 \rho^2 ]
   +\frac{\lb \rho}{12} K_1(\rho \lb) [11+ \frac{5}{4} \lb^2 \rho^2]
\ea
  The calculated $\varphi(s,t)$ (\ref{fei2}) is an eikonal analog with taking account
 of the hadron form-factor
 of the expression  obtained by West and Yennie \ci{wy}
 from the Feynman diagram.

    If we take the Born term of the hadron amplitude in the Regge form
\ba
   F^B_h(s,q) = h \ \exp{(-B(s) q^2)},
\ea
 and only the first term of (\ref{ce1}) that corresponds to
 the Coulomb amplitude without the form-factor,
 we obtain the ordinary representation  for the total phase factor
in the region of small transfer momenta
\ba
  \varphi \simeq \ln{\frac{B(s) q^2}{2}} + \gamma .
\ea

  Our eikonal representation for the $\varphi(s,t)$ (\ref{fei2})
 is valid in a wide region of $t$. If we take the correct
 hadron scattering eikonal that describes the experimental differential
 cross section including the domain of the diffraction dip,
 we can calculate the $\varphi(s,t)$     for that region
 of $t$.  Note, that the calculated term has a real and
 a non-small imaginary part.
    The phase factor can be very important for the calculation of  spin
 correlation parameters with taking into account
  the electromagnetic-hadron interference effects
 in the domain of the diffraction dip. In this region, the imaginary part
 of the hadron scattering amplitude has a zero and these effects can be not small.
 Beforehand, we don't know how large the additional contributions
  will be and
 how the phase factor affects on  magnitudes
 of the spin correlation parameters.
  The largest effects will be in the energy range
  where the diffraction minimum has a strictly defined form and, thus,
  the real part of the non-flip hadron amplitude will be small.

     We can take, for example,  the eikonal of the spin-non-flip hadron
  amplitude
  obtained in \ci{zpc} that provides
  a quantitative description of high-energy  experimental data:
\ba
 -\chi_h(s, b)=h_1(s) \exp{(-\mu(s) \sqrt{b^2+b_{p}^2(s)})}
                     [1-h_2(s) \exp{({-\mu(s) \sqrt{b^2+b_{p}^2(s)}})}].
			    \lab{fh}
\ea
 where $b_{p}(s)$ and $\mu(s)$
 are the effective radius and mass, and
  calculate the coulomb-hadron phase-factor \cite{selph}.

  As this model is based on the crossing symmetry of the scattering amplitude,
 the parameters in this eikonal are complex and, hence,
 the obtained scattering amplitude has a real and an imaginary part.
 Using this eikonal we can calculate  the total phase factor,
 for example, for $\sqrt{s} =  50 \  GeV$
  (the first starting energy of RHIC).
 Note that our results depend weakly on the model
 for the hadron spin-non-flip  eikonal.
 Different models should give the same
 $\sigma_{tot}, \rho(s)= ReF(s,0)/ImF(s,0)$ and
 differential cross sections in a wide region of
 energies and momenta transfer.

  The value of the coulomb-hadron phase-factor in the domain of the
  diffraction dip is large in magnitude
 with  a positive maximum
 of the real part at $-t=1.25 \ GeV^2$
 and  a negative maximum.
 at $-t=1.45 \ GeV^2$.
The imaginary part has one negative
 maximum
 at the point of the minimum of  the diffraction dip.

  Let us calculate, using the obtained $\varphi(s,t)$,
    an additional contribution to  the analyzing power $A_N$
   and double spin correlation parameter $A_{NN}$
    owing to the electromagnetic-hadron interference $A_{N}^{emN}$
   and $A_{NN}^{emN}$
   in the diffraction dip domain of the proton-proton elastic scattering.
  This contributions should be taken into account
  when we want to know the true size
  of the contribution  to
  $A_{N}$ and  $A_{NN}$
  due to the hadron spin-flip amplitude.
  Let us again take the hadron spin-non-flip amplitude
  calculated in \cite{zpc}
  and calculate the size of $A_{N}^{emN}$ in the diffraction dip domain
  for example, at energy HERA $\sqrt{s} = 40 \ GeV$.
     The calculated analyzing power is
 shown in Fig.3a.
  Its form is defined by the
  form of the diffraction minimum and the behavior of the imaginary
  part of the hadron spin-non-flip amplitude.
  As it is noted above, our result  coincides qualitatively with
  the calculation  of \ci{bs} up to the moment
  when the imaginary part of the hadron
  non-flip amplitude changes in sign and the polarization
  has a negative maximum.
  On the whole,
   the picture is similar to that of behavior of $\varphi$.
 The analysis shows that the impact of  $\varphi$ leads
  in the proton-proton scattering to the
 increase of the analyzing power approximately by $ 50\% $ of the effect.
 Hence,
 if only this impact were
  negative,  the size of $A_N^{emN}$ becomes very small.

   Now let us examine such an additional contribution to the double spin
   correlation parameter.
   It is clear that the large contribution  to $A_{NN}^{emN}$
 comes from the interference of electromagnetic amplitudes
  $ \phi^{em}_1$ and $\phi^{em}_2$.
 but in our case it
            is cancelled  completely by
 the contribution of  $ 2  |\phi^{em}_{5} |^2 )$.

 So, we have  the magnitude of $A_{NN}^{emN}$
  both dependent on the real part of the hadron spin-non-flip amplitude and
  $\phi_2^{em}$.
    Our calculation of $A_{NN}^{emN}$ is shown in Fig. 3b.

\begin{figure}
\begin{flushleft}
\mbox{\epsfysize=50mm\epsffile{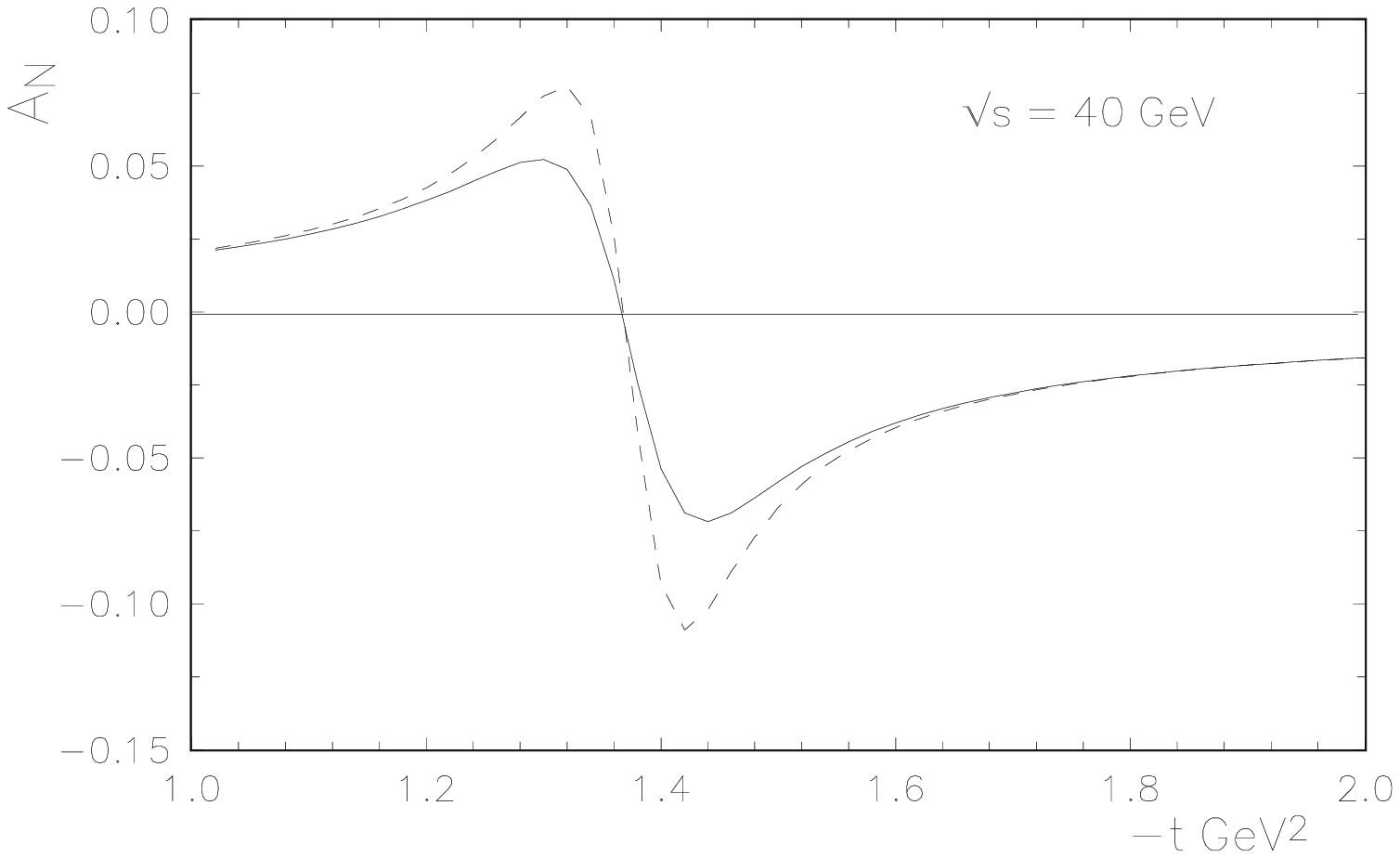}}
\end{flushleft}
\vspace{-5.7cm}
\begin{flushright}
\mbox{\epsfysize=50mm\epsffile{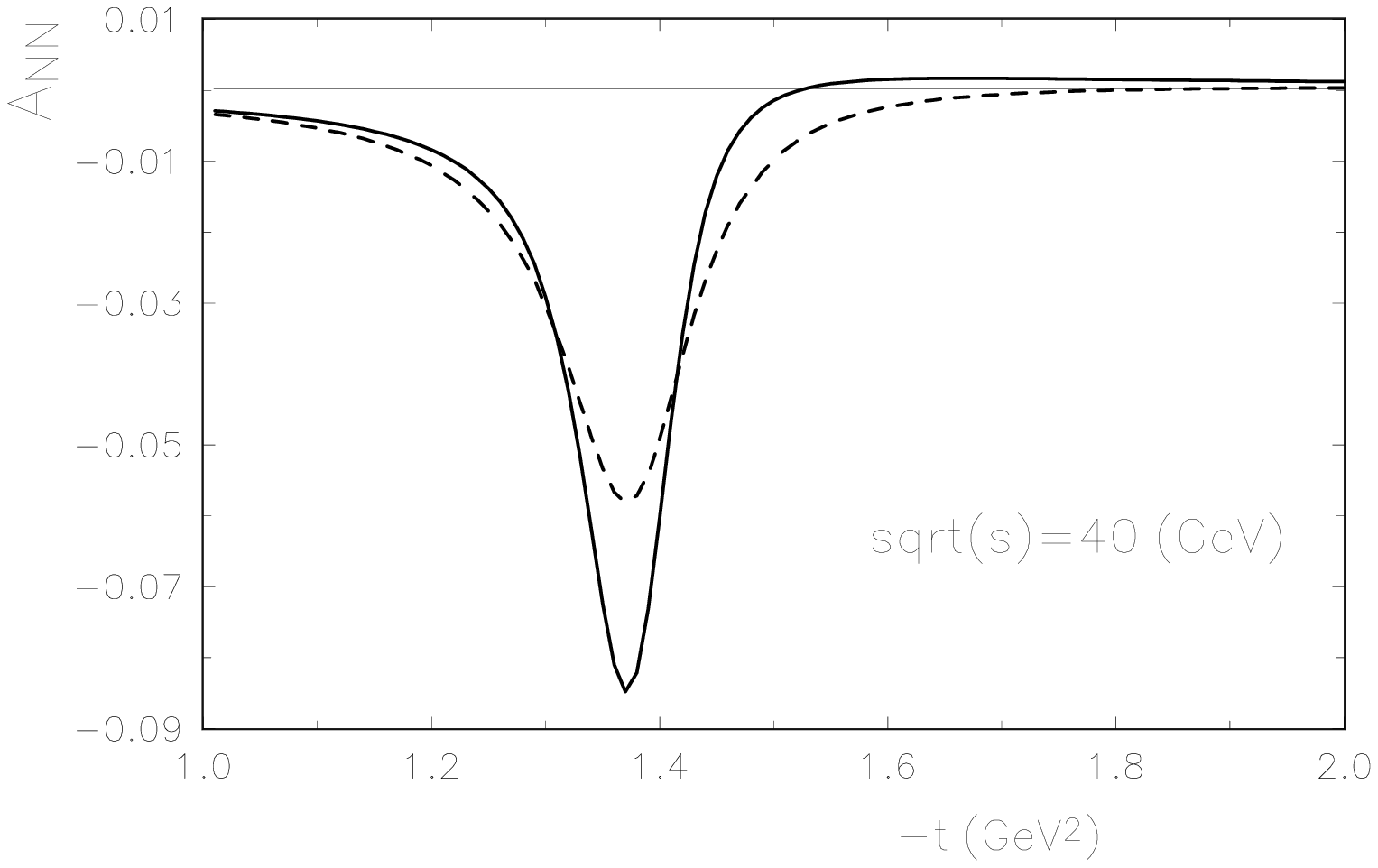}}
\end{flushright}
\centering
 {\bf a) \hspace{7.cm} b)}
\caption{ The calculated a) $A_{N}^{emN}$ and b) $A_{NN}^{emN}$ }
\label{F4}
\end{figure}

  We obtain the negative non-small contribution which
  reduces   the size of the double spin correlation parameter
   owing to the hadron-spin-flip amplitude.
  Of course, the form and size of $A_{NN}^{emN}$
  are mostly defined   by the form and size of the diffraction minimum.

   In spite of the large  contribution of the hadron-spin-flip
 amplitude we can see that taking account of the electromagnetic
  amplitudes leads to visible changes in spin correlation effects.
 Note that in our case
 the part of the contribution of the interference
 electromagnetic and hadron-spin-non flip amplitude is reduced by  terms
  of the interference between of the electromagnetic and hadron-spin-flip
  amplitude.  On the whole, the spin correlation effects
  somewhat decrease
  and  slightly change the positions of maxima.


  Using the eikonal  representation
 we
 obtain the terms corresponding to the electromagnetic-hadron interference
 in the second Born approximation
 taking into account the form-factors of interacting nucleons
 in a wide region of transfer momenta
 up to the diffraction dip domain.
 It allows us to calculate an additional contribution to the
 analyzing power $A_{N}$ and double spin correlation parameter
 $A_{NN}$ owing to the interference of the electromagnetic
 and hadron-spin-non-flip  amplitudes.
 As a result, we
 obtain, in the domain of the diffraction minimum,
 not small spin correlation effects due to the interference of
 the spin-non-flip elastic scattering amplitude and the
  electromagnetic spin-flip
 amplitude.
   The obtained energy dependence   of
  $A^{emN}_{N}$  and   $A^{emN}_{NN}$ shows that these contributions have
  to be taken into account when we extract the size and
   energy dependence of
   the hadron spin-flip  amplitude
   from the experimental data on  $A_{N}$ and $A_{NN}$
    at high energies  up to the top of RHIC energy.

\end{document}